\documentclass[preprint,aps,prl,showpacs]{revtex4}
\usepackage{graphicx}
\begin{document}

\title{Shape of the spatial mode function of photons generated in noncollinear spontaneous parametric downconversion}

\author{Gabriel Molina-Terriza,$^{1}$ Stefano Minardi,$^{1,*}$ Yana Deyanova,$^{1,2}$ Clara I. Osorio,$^{1,2}$ Martin Hendrych,$^{1,2}$ and Juan P. Torres$^{1,2}$}

\address{$^1$ICFO-Institut de Ciencies Fotoniques, Universitat Politecnica de Catalunya, 08034 Barcelona, Spain}

\address{$^2$Department of
Signal Theory and Communications, Universitat Politecnica de Catalunya, 08034 Barcelona, Spain}

\begin{abstract}
We show experimentally how noncollinear geometries in spontaneous parametric downconversion induce ellipticity of the shape of the spatial mode function. The degree of ellipticity depends on the pump beam width, especially for highly focused beams. We also discuss the ellipticity induced by the spectrum of the pump beam.
\end{abstract}


\maketitle

\newpage
Spontaneous parametric downconversion (SPDC), namely the generation of two lower-frequency photons when a strong pump interacts with a nonlinear crystal, is a reliable source for generating pairs of photons with entangled properties. The photons generated in SPDC exhibit a rich structure, i.e., the two-photon quantum state is described by its polarization, spatial shape and frequency spectrum. However, the corresponding entangled states generally take advantage of only a portion of the total two-photon quantum state.

To date most applications of parametric downconversion in quantum systems make use of polarization entanglement as the quantum resource \cite{zeilinger1,nielsen1}. Such entanglement is confined to a two-dimensional Hilbert space. On the contrary, both frequency entanglement and spatial entanglement occur in an infinite-dimensional Hilbert space \cite{law1,torres1}, which opens a wealth of opportunities to enhance the potentiality of quantum techniques.

Here we are interested, in particular, in the spatial properties of the two-photon state embedded in the corresponding two-photon amplitude, or spatial mode function. The multidimensional entangled states, or qudits, can be encoded in orbital angular momentum (OAM) 
\cite{mair1,arnaut}. This provides infinite-dimensional alphabets \cite{molina1} that can be used to conduct proof-of-principle demonstrations of quantum protocols requiring higher-dimensional Hilbert spaces for its implementation. Illustrative examples include the violation of Bell's inequalities with qutrits \cite{vaziri1} and the implementation of a quantum coin tossing protocol \cite{molina2}.

Most previous investigations of the spatial shape of entangled photons addressed nearly collinear phase-matching geometries in which pump, signal and idler photons propagate nearly along the same direction, and where the experimental conditions safely allow to neglect the Poynting vector walk-off between the interacting waves. Such is the case that holds approximately in most of the observations reported to date (see, in particular, Refs \cite{mair1,walborn1,howell1}). Under these conditions, the detailed structure of the corresponding quantum states has been experimentally elucidated \cite{molina3}.

Actually, in genuine noncollinear geometries, i.e. in situations where there is a nonzero angle between the directions of propagation of the downconverted photons and the pump beam, both the spin angular momentum \cite{migdall1} and the spatial shape of the entangled photons depend strongly on the propagation direction of the photons. When strongly focused pump beams are used, the noncollinear effects can be made clearly visible, as observed recently by Altman and co-workers \cite{barbosa1}. The impact of such effects can be made important even when only purely geometrical features are considered \cite{molina4}, and can become dominant in the case of highly noncollinear settings such as transverse-emitting configurations \cite{torres2}. It has been theoretically shown that such noncollinear geometries result in the generation of asymmetrical spatial shapes of the two-photon mode function. The analysis of the ellipticity of the spatial shape is of paramount importance for implementing quantum information protocols based on spatially encoded information \cite{vaziri1,molina2}, for the study of entanglement of continuous variables using the momentum and position of two-photon states \cite{howell1}, as well as for those applications addressed to quantum imaging \cite{pittman1,soutoribeiro1}.

Quantum information protocols implemented up to date can be described by a mode function which is paraxial in a suitable transverse frame centered of the central signal and idler wave vectors. The observation of the spatial shape of such mode function, which is of interest in this letter, is a step forward in the development of new quantum information protocols based on spatial information. On the other hand, one can also consider the global mode function that describes the two-photon state \cite{arnaut}, which is relevant for the elucidation of angular momentum balance in SPDC \cite{bloembergen1}.

The aim of this paper is to show experimentally that the spatial shape of the downconverted photons exhibits ellipticity that depends on the noncollinear geometry used. In this sense, we show that the ellipticity can be tailored by a suitable choice of the different experimental parameters. In particular, for large pump-beam sizes the spatial mode function holds approximately the symmetry of the pump beam shape. Furthermore, we show that the observed ellipticity can also be influenced by the frequency bandwidth of the pump beam, when the bandwidth of the interference filters located in front of the detectors allows for it.

Let us consider a quadratic nonlinear optical crystal of length $L$, which is illuminated by a laser pump beam propagating in the $z$ direction. We assume that there is no Poynting vector walk-off between the interacting waves. The amplitude of the paraxial pump beam, which is treated classically, writes $E_p({\bf x},z,t)=\int d\omega_p\, d {\bf P} E_0 \left(\omega_p\, ,{\bf P} \right) \exp \left[ i k_p z + i {\bf P} \cdot {\bf x}-i\omega_p t \right]+h.c.$ where ${\bf x}=(x,y)$ is the position in the transverse plane, ${\bf P}$ is the transverse wavenumber, $\omega_p$ is the angular frequency of the pump beam, $k_p=\sqrt{(\omega_p n_p/c)^2-|{\bf P}|^2}$ is the longitudinal wave number inside the crystal, $n_p$ is the refractive index at the central pump angular frequency, and $E_0(\omega_p\, ,{\bf P})$ is the amplitude of the pump beam.

The downconverted photons propagate in the $yz$ plane. The two-photon quantum state at the output of the nonlinear crystal, within the first order perturbation theory, can be written as $|\Psi>=\int d\omega_s\, d\omega_i\, d {\bf p}\, d{\bf q} \,\Phi \left(\omega_s,\omega_i,{\bf p},{\bf q} \right) a_s^{\dagger} \left( \omega_s, {\bf p}) a_i^{\dagger}(\omega_i,{\bf q} \right) |0,0>$, where ${\bf p}=(p_x,p_y)$ is the transverse wavenumber of the signal photon \cite{molina4}, ${\bf q}=(q_x,q_y)$ is the corresponding transverse wavenumber of the idler photon, and $\omega_s$ and $\omega_i$ are the angular frequencies of the signal and idler photons, respectively. The two-photon amplitude, or spatial mode function $\Phi$ then writes \cite{torres2,joobeur1}
\begin{equation}
\label{central1} \Phi \left(\omega_s,\omega_i, {\bf p},{\bf q} \right) = E_0 \left(\omega_s+\omega_i, p_x+q_x, \Delta_0 \right)
\mbox{sinc} \left( \Delta_k L/2 \right) \exp \left( -i\Delta_k L/2 \right),
\end{equation}
with $\Delta_k= k_p - k_s \cos \varphi_1-k_i \cos \varphi_2 - p_y \sin \varphi_1 -q_y \sin \varphi_2$ which comes from the phase matching condition in the longitudinal direction $z$, and $\Delta_0=p_y \cos \varphi_1+ q_y \cos \varphi_2 -k_s \sin \varphi_1 -k_i \sin \varphi_2$, which comes from the phase matching condition in the transverse dimension $y$. The angles $\varphi_1$ and $\varphi_2$ are the internal angles of emission of the signal and idler photons, respectively, $k_s=\sqrt{(\omega_s n_s/c)^2-|{\bf p}|^2}$ is the longitudinal wavenumber of the signal photon, and $k_i=\sqrt{(\omega_i n_i/c)^2-|{\bf q}|^2}$ the corresponding one for the idler photon. $n_s$ and $n_i$ are the refractive indexes at the central frequencies of the signal and idler photons. The longitudinal wavenumber of the pump beam writes $k_p=\sqrt{(\omega_p n_p/c)^2-(p_x+q_x)^2- \Delta_0^2}$, with $\omega_p=\omega_s+\omega_i$. Under the thin crystal approximation, if we assume a monochromatic pump beam and signal and idler photons, the mode function given in Eq. (\ref{central1}) writes \cite{molina4} $\Phi \left( {\bf p},{\bf q} \right) \propto E_0 \left( p_x+q_x,p_y \cos \varphi_1+q_y \cos \varphi_2 \right)$. Notice that in this case the two-photon mode function is proportional to the wavevector distribution of the pump beam, evaluated at the vectorial sum of the signal and idler wavevectors.

In our experiment the downconverted photons emitted by the nonlinear crystal are conveyed into 2-$f$ optical systems, which provide an image of the spatial shape of the two-photon state $\Phi({\bf p},{\bf q})$ \cite{abouraddy1}. If we consider a monochromatic pump beam, very narrow interference filters in front of the detectors, and ideal point-like detectors, the coincidence rate $I({\bf x}_1^0, {\bf x}_2^0)$, which is proportional to the probability of detecting in coincidence a signal photon at ${\bf x}_1^0$ and an idler photon at ${\bf x}_2^0$, writes
\begin{equation}
\label{shape}
I({\bf x}_1^0, {\bf x}_2^0)=\left|\Phi\left(  \frac{2 \pi{\bf x}_1^0}{\lambda_s^0 f},\frac{2 \pi{\bf x}_2^0}{\lambda_s^0
f}\right)\right|^2,
\end{equation}
where $\lambda_s^0$ and $\lambda_i^0$ are the central wavelengths of the signal and idler photons, respectively.

\begin{figure}
\includegraphics[width=.9\columnwidth]{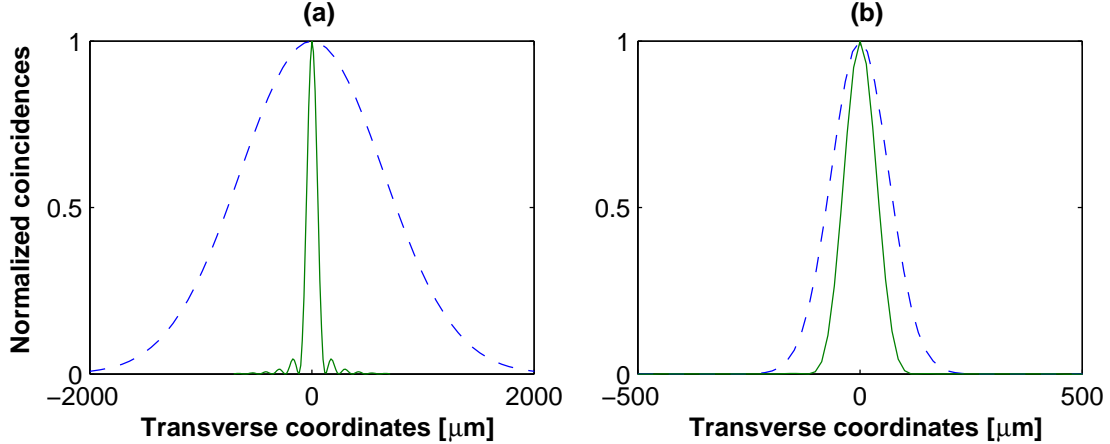}
\caption{ Calculated coincidence rate for two different pump-beam widths ($w_0$): (a) $w_0=50\mu $m, and (b) $w_0=500 \mu$m. Solid lines: $y$ coordinate ({\em horizontal cut}); dashed lines: $x$ coordinate ({\em vertical cut}). Length of the LiIO$_3$ crystal: $L=5$mm. Internal angle of emission of the downconverted photons: $\varphi_1 \simeq 17.1^o$.}
\end{figure}

In Fig. 1 we plot the coincidence rate for two different values of the pump-beam width. Both cases display a different degree of spatial ellipticity. From Eqs. (\ref{central1}) and (\ref{shape}), it follows that while the vertical width of the correlation function only depends on the spatial bandwidth of the pump beam, the horizontal one is further constrained by the phase matching conditions of the crystal. The important parameter that controls the degree of spatial ellipticity induced by the noncollinear geometry is the noncollinear length \cite{torres3}, which writes $L_{nc}=w_0/\sin \varphi_1$ with $w_0$ being the pump width. For small $w_0$ or large noncollinear angles $\varphi_1$ such that the noncollinear length is smaller than the crystal length ($L_{nc} < L$), the spatial shape of the signal photon is highly elliptical. This is the case represented in Fig. 1(a). For larger pump-beam widths, the noncollinear length is increasingly larger so that the ellipticity is correspondingly suppressed, as can be observed in Fig. 1(b). This is the working experimental regime for the most of the experiments performed up to date that implement protocols based on spatial quantum information \cite{vaziri1,molina2}.

The effects described here are due to the noncollinear propagation directions of the photons emerging from the downconverting crystal. The spatial ellipticity induced by the noncollinear configuration is thus enhanced if the angles of propagation of the photons ($\varphi_1$) are large, or if the width in transverse wavevector space ($1/w_0$) of the downconverted photons is also large. The importance of the simultaneous effect of both quantities is naturally revealed through the noncollinear length.

\begin{figure}
\includegraphics[width=.9\columnwidth]{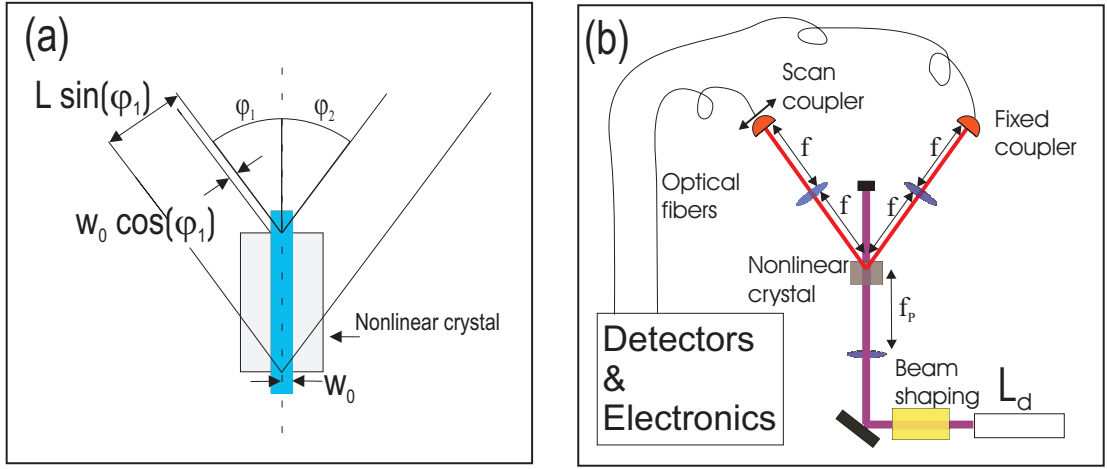}
\caption{Sketch of the experimental configuration. (a) Noncollinear configuration, the relevant parameters are shown as they are inside the crystal, for values outside the crystal the refraction has to be taken into account. (b) Experimental setup. The light from a $405$nm diode laser with 0.6nm bandwidth is passed through a spatial filter and focused into the LiIO$_3$ crystal. The downconverted photons are coupled into multi-mode fibers after traversing through 2-$f$ systems.}
\end{figure}

We set up an experiment (see Fig. 2), to show the spatial properties of the photons in a noncollinear configuration. We use a type-I degenerate noncollinear SPDC in a $L=5$mm thick lithium iodate (LiIO$_3$) crystal.  The crystal is cut in a configuration such that neither of the interacting waves exhibit a Poynting vector walk-off. This is an important point to consider when choosing the most appropriate experimental configuration. For highly focused pump beams, where noncollinear effects are to be clearly observable, modifications of the spatial two-photon amplitude induced by the Poynting vector walk-off are no longer negligible \cite{torres3}. The internal angle of emission of the downconverted photons is $\varphi_1 = -\varphi_1 =\simeq 17.1^o$, the largest that can be achieved with this crystal \cite{handbook}.

The pump source is a multimode continuous-wave diode laser emitting light centered at a wavelength of $\lambda_p^0=405$nm with an estimated bandwidth of around $0.6$nm. The spatial mode at the output of the diode laser is spatially filtered in order to obtain an approximate Gaussian beam with a beam-waist radius of about $500\mu$m and power up to 25mW. Lenses of different focal length ($f_p$) are placed before the crystal to control the input pump-beam waist.

\begin{figure}
\includegraphics[width=.9\columnwidth]{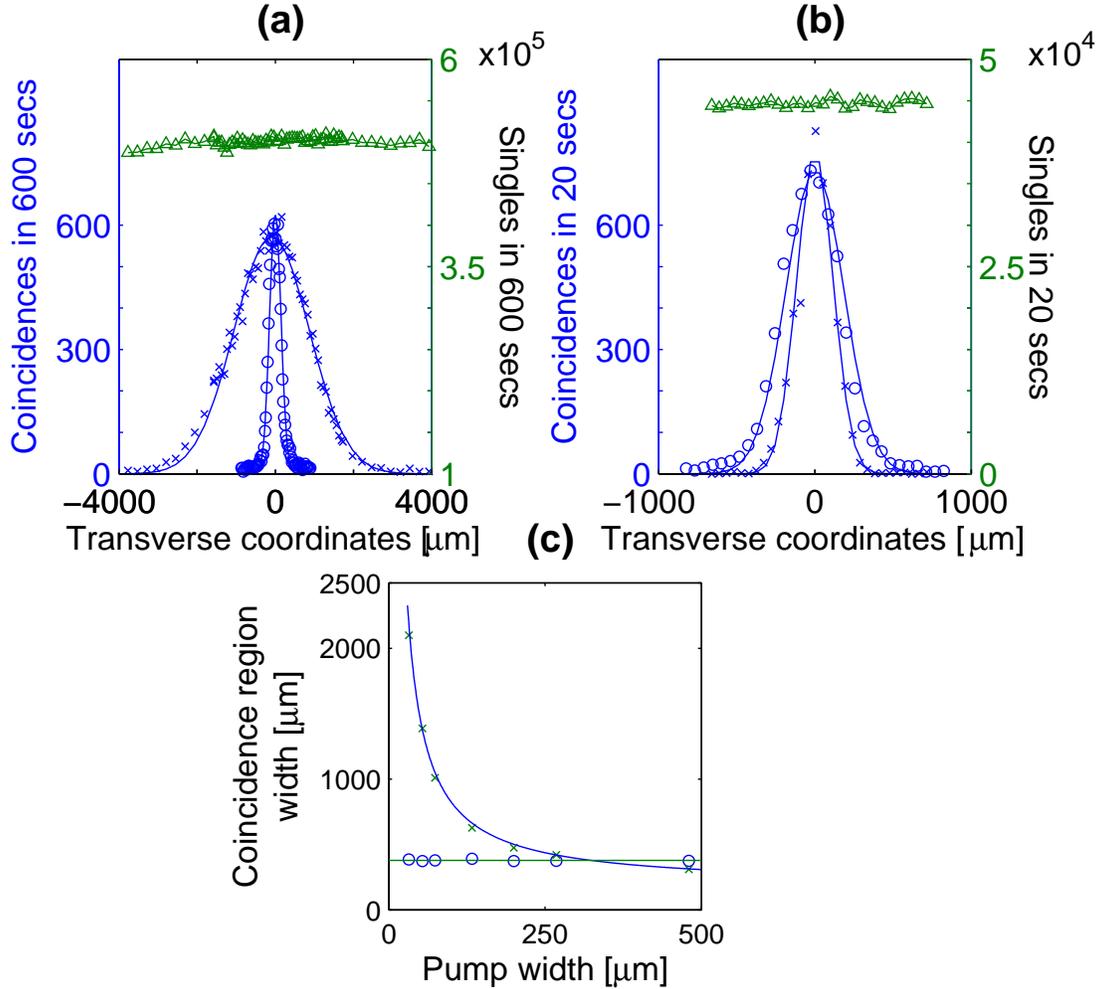}
\caption{Experimentally measured coincidence rate with a pump-beam width (a) $w_0\simeq32 \mu $m, and (b) $w_0\simeq500 \mu$m. (c) Width (half width, $1/e^2$) of the conditional coincidence rate along the two transverse directions, $x$ and $y$, as a function of the pump-beam width. Circles: $y$ dimension; crosses: $x$ dimension; triangles: singles along the $x$ dimension. Solid lines are the best fit to the experimental data. All other experimental conditions as described in Fig. 1.}
\end{figure}

As depicted in Fig. 2(b), after the crystal a 2-$f$ system of $250$mm focal length is used for each of the downconverted beams. The  photons are then coupled into multimode optical fibers and sent to the single-photon detectors. The coupling lenses are mounted on a XY translation stage. The one corresponding to the signal-photon beam has been equipped with stepper motors that allow for scanning in the transverse plane. The coupler in the idler beam was used as a fixed reference. Therefore, the measurements obtained correspond to $I \left( {\bf x}_1^0,{\bf x}_2^0=0 \right)$. Broadband colored filters  in front of the couplers are used to remove scattered radiation at 405nm. Pinholes with diameters of $100\mu$m and $150\mu$m (depending on the experiment) are attached to the couplers in order to increase the resolution.

Fig. 3 presents the main results of our experiment. We plot the coincidence rate, which corresponds to scanning the signal coupler position along two different orthogonal directions, while keeping fixed the position of the idler coupler. In the $x$ transverse dimension ({\em vertical cut}), the spatial shape can well be described within the thin crystal approximation with a Gaussian shape with a beam width of $w_{x}=\lambda_s f/(\pi w_0)$. In Fig. 3(a), the pump-beam width, $w_0=32\mu m$, yields a noncollinear length of $L_{nc} \simeq 100\mu$m, which is much smaller than the crystal length.  Fig. 3(b) shows the coincidence rate for a pump-beam width of $w_0\simeq 500\mu$m. For $w_0=32 \mu m$ (Fig. 3(a)), the measured width of the two-photon amplitude in the $x$ dimension is $w_x\simeq 960 \mu m$, while in the $y$ dimension, we measured $w_y \simeq 150 \mu m$. For $w_0=500 \mu m$, the corresponding measurements yield $w_x \simeq 120 \mu m$ and $w_y \simeq 180 \mu m$ (Fig. 3(b)). The ellipticity of the state is clearly reduced for a larger pump beam width. We have experimentally verified that the use of narrow band filters ($10$nm bandwidth, FWHM) in front of the detectors does not modify the measured shape of the coincidence rate significantly, although it modifies the single counts spatial shape.

Fig. 3(c) shows the dependence of the width (half width, $1/e^2$) of the coincidence rate as a function of the pump-beam width. It shows that the state ellipticity is clearly present when a strongly focused pump beam is used. Small pump-beam widths correspond to small noncollinear lengths when compared to the crystal length. With a monochromatic pump and large pump beams we expect the $x$ and $y$ widths to be equal. Nevertheless, a  finite spectrum of the pump beam strongly affects the width in  the $y$ direction.

The use of a non-monochromatic pump beam (multimode beam), combined with the use of wide enough interference filters in front of the photon detectors (or even no interference filters at all), can in fact induce ellipticity of the spatial two-photon amplitude. In this case, the coincidence rate is given by
\begin{equation}
\label{rate1} I \left( {\bf x}_1^0,{\bf x}_2^0 \right) = \int_B
d\omega_s\, d\omega_i\, \int_{\Omega} d {\bf x}_1\, d{\bf x}_2 \,
\left| \Phi \left(\omega_s,\omega_i,\frac{2 \pi{\bf
x}_1}{\lambda_s^0 f},\frac{{2 \pi \bf x}_2}{\lambda_i^0 f}
\right)\right|^2\left| H_s(\omega_s) \right|^2 \left|
H_i(\omega_i) \right|^2,
\end{equation}
where $H_s(\omega_s)$ and $H_i(\omega_i)$ are the corresponding interference filters located in front of the detectors, $\Omega$ is the finite area of the pinholes centered around ${\bf x}_1^0$ and ${\bf x}_2^0$, respectively, and $B$ designates the corresponding bandwidth of the signal, idler and pump waves.

\begin{figure}
\includegraphics[width=.9\columnwidth]{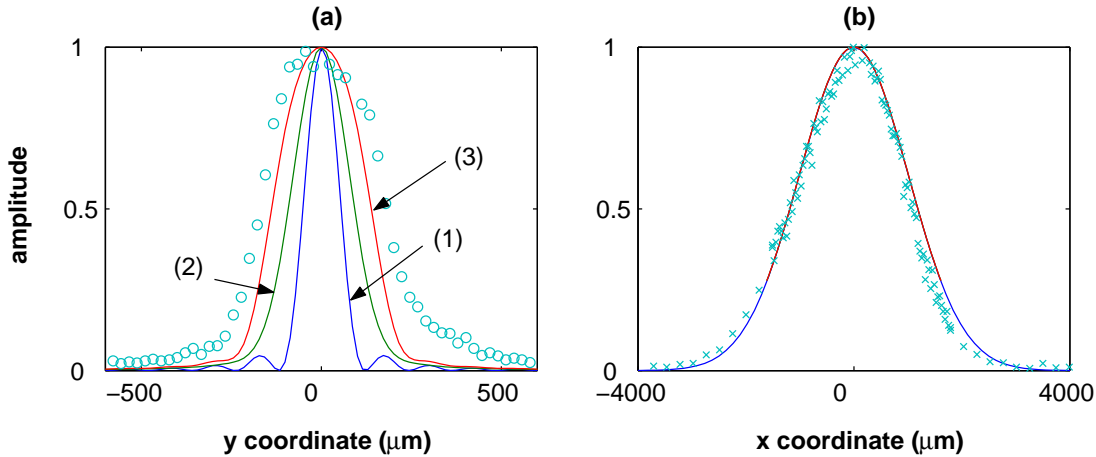}
\caption{Coincidence rate for different pump bandwidths. (a) $y$ transverse dimension and (b) $x$ transverse dimension. The multimode pump beam is approximated by a Gaussian centered at $405$nm. The Gaussian bandwidth (FWHM) is $0.6$nm (curve $3$) and  $0.3$nm (curve $2$). For comparison, we also plot the case for a monochromatic pump beam (curve $1$). The circles and crosses corresponds to experimental data, as shown in Fig. 3(a).  In (b), all curves coincide. The pump beam width is $w_0=32 \mu$m. All other parameters as described in Fig. 1.}
\end{figure}

In Fig. 4 we plot the coincidence rate along the $x$ and $y$ transverse dimensions, when considering a multimode pump beam and Gaussian interference filters ($\Delta \lambda=10$nm bandwidth, FWHM). We can see that the spectrum of the pump beam modifies the spatial two-photon amplitude in the $y$ transverse dimension ({\em horizontal cut}). In the $x$  dimension, the spectrum of the pump beam does not modify the spatial shape. The different influence of the pump spectrum on the spatial shape in both transverse dimensions is a clear indication of the importance of the noncollinear configuration for observing this frequency induced spatial ellipticity. Indeed, for large pump beams the state ellipticity can be reversed, so that the width in the $x$ dimension is smaller than in the $y$ dimension, as can be seen in Figure 3(c).

In conclusion, we have shown experimentally that the SPDC noncollinear configurations induce ellipticity of the spatial two-photon amplitude. The noncollinear length measures the importance of the noncollinear effects in the shaping of the two-photon amplitude of the two-photon state. This turns out to be of great importance when designing and implementing quantum information protocols, especially when using highly focused beams. Furthermore, we have shown that the spectrum of the pump beam, when the bandwidth of the interference filters located in front of the detectors is wide enough, can also induce ellipticity of the spatial two-photon amplitude.

This work was supported by the Generalitat de Catalunya and by
grants BFM2002-2861 and FIS2004-03556 from the Government of Spain.

$^*$ S. Minardi is currently with the Department of Electronics, T. E. I. Crete at Chania, Romanou 2, 73100 Chania, Greece.

\end{document}